\documentclass[aps,prl,twocolumn,showpacs]{revtex4-1}
\usepackage{graphicx}

\begin{document}

\title{GHz Operation of Nanometer-Scale Metallic Memristors: Highly Transparent Conductance Channels in Ag$_{2}$S Devices}

\author{A. Geresdi}
\author{M. Csontos}
\email{csontos@dept.phy.bme.hu}
\author{A. Gubicza}
\author{A. Halbritter}
\author{G. Mih\'aly}
\affiliation{Department of Physics, Budapest University of Technology and Economics and \\
Condensed Matter Research Group of the Hungarian Academy of Sciences, 1111 Budapest, Budafoki ut 8, Hungary}

\date{\today}

\begin{abstract}
The nonlinear transport properties of nanometer-scale junctions formed between an inert metallic tip and an Ag film covered by a thin Ag$_{2}$S layer are
investigated. Suitably prepared samples exhibit memristive behavior with technologically optimal ON and OFF state resistances yielding to resistive switching
on the nanosecond time scale. Utilizing point contact Andreev reflection spectroscopy we studied the nature of electron transport in the active volume of the
memristive junctions showing that both the ON and OFF states correspond to truly nanometer scale, highly transparent metallic channels. Our results demonstrate
the merits of Ag$_{2}$S nanojunctions as nanometer-scale memory cells with GHz operation frequencies.
\end{abstract}


\maketitle

The ongoing miniaturization beyond the limitations of nowadays CMOS technology is a major challenge in nanosciences \cite{itrs2011,Chau2007,6200837}. Using
individual atoms or molecules in nanoelectronic circuits has been a breakthrough towards the ultimate single atomic size limit \cite{Heath2003,Schirm2013}. The
persisting technological difficulties in the reliable assembly of low resistance single molecule devices, however, still represent a major barrier to fast
applications preferring low $RC$ time constants with the capacitance of the environment. Alternatively, reversible solid state electrochemical reactions have
been proposed to form tunable atomic scale junctions between metallic electrodes. The first results are extremely promising for the short term realization of
highly integrated information storage applications \cite{Yang2013,nature03190,nature06932,nmat2023,advmat2007,IEEE2010,nmat2748,Torrezan2011}. The resistive
state of such a memory element, called \emph{memristor} \cite{nature06932,nmat2023,ADMA:ADMA200900375,1083337,doi:10.1080/00018732.2010.544961,Yang2013}, is
altered by biasing the device above its writing threshold $V_{\rm th}$. Readout is performed at lower signal levels which preserve the stored information.

Chalcogenide compounds have been put forward in the context of the ``atomic switch'' \cite{nature03190}, consisting of an inert metallic electrode (Me) and a
Ag layer capped with the solid state ionic conductor Ag$_{2}$S. Upon positively biasing the Ag electrode with respect to Me, a metallic Ag propulsion is grown
on the Ag$_{2}$S-Me interface shunting the electrodes thus creating the non-volatile ON state of the device
\cite{terabe:10110,0957-4484-20-9-095710,C0NR00298D,Masis2011,C0NR00951B}. Real-time high resolution transmission electron microscopy (HRTEM) imaging
\cite{doi:10.1021/nn100483a} and first principle band structure calculations \cite{nl0711054,wang:152106} suggested that a structural phase transition in the
Ag$_2$S layer also plays a role in the resistive transition \cite{wagenaar:014302}.

Since the pioneering experiments reported on Ag-Ag$_{2}$S-Pt devices \cite{nature03190} the development of memory cells based on memristive systems has
achieved a remarkable progress. Beside providing an interesting model system for neural networks \cite{ADMA:ADMA200903680,Ohno2011,Pickett2013} Ag$_{2}$S based
devices have been utilized as nanometer-scale non-volatile memory elements \cite{nature03190,C0NR00951B}. However, the best performing Ag$_{2}$S devices
\cite{nature03190} have been operated only up to $\sim$10~MHz frequencies presumably due to the typically $\geq$100~k$\Omega$ OFF state resistances which
require the monitoring of technically unfavorable low currents and give rise to larger $RC$ time constants which are inconvenient for GHz applications
\cite{jz900375a}. In a tantalum oxide based system sub-nanosecond switching times were shown \cite{Torrezan2011} in a significantly larger, lithographically
defined structure whose operation relies on the reconfiguration of oxygen vacancies. Here we demonstrate devices which not only approach the atomic size limit
but are also capable of GHz operation.

\begin{figure}[t!]
\includegraphics[width=\columnwidth]{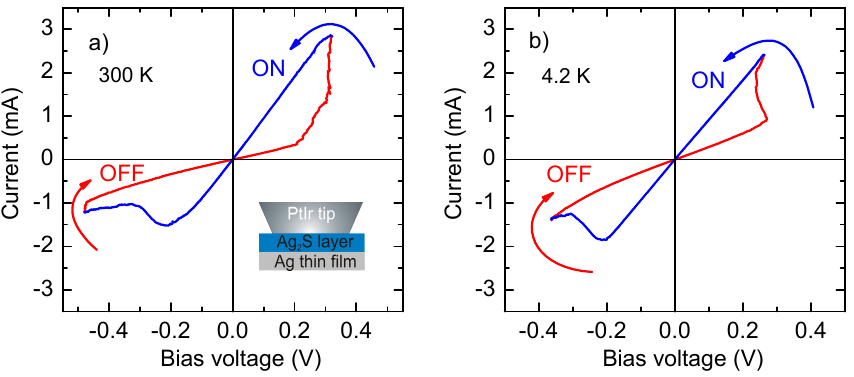}
\caption{(Color online) Typical I-V characteristics of Ag$_2$S based memristive junctions at room temperature (a) and at 4.2 K (b). Positive voltage
corresponds to the positive biasing of the Ag layer with respect to the PtIr electrode. The arrows indicate the direction of the bias sweep. The inset shows
the schematic cross section of the sample.} \label{Fig1}
\end{figure}

We study resistive switching in voltage biased nanojunctions created between inert metallic tips and 10--100 nm thick Ag$_2$S surface layers deposited on Ag
thin film samples by using an STM setup as illustrated schematically in the inset of Fig.~\ref{Fig1}(a). While longer sulfur deposition times resulted in
semiconducting Ag$_2$S layers in agreement with previous reports \cite{0957-4484-20-9-095710,C0NR00298D,Masis2011}, below an approximate thickness of 20~nm
metallic conductance with technologically optimal device resistances was found over the wide temperature range of 4.2--300~K \cite{GeresdiMRS2011} \emph{both}
in the ON and OFF states. These characteristics enable GHz operation. Here we focus on switching phenomena observed in such all-metallic junctions. We use a
superconducting Nb tip to study the nonlinear differential conductance on the voltage scale of the superconducting gap, $eV\leq\Delta=1.4$~meV~$\ll V_{\rm
th}$. By utilizing the theory of charge conversion at the interface of a normal metal and a superconductor
\cite{Andreev1964,PhysRevB.25.4515,PhysRevB.54.7366,Soulen1998} we quantitatively evaluate the reconfiguration of the conducting channels in the nanojunction
and thus demonstrate that resistive switching takes place in highly transparent devices with an effective junction area of 2--5 nm in diameter.

Numerous nanoscale contacts with reproducible I-V characteristics were created by gently touching the sample surface with a mechanically sharpened PtIr or Nb
tip. The ON and OFF state resistances $R_{\rm ON}$ and $R_{\rm OFF}$ were probed in a narrow voltage window of $\pm50$~mV $\ll V_{\rm th}$. For more
experimental details see the Appendix and Refs.~\citenum{GeresdiMRS2011, Kotai1994588}.

A typical room temperature I-V trace is shown in Fig.~\ref{Fig1}(a) for a PtIr tip. Applying an increasing positive voltage on the Ag electrode in the high
resistance ($R_{\rm OFF}\approx 0.5$~k$\Omega$) state first a linear current-voltage dependence is observed. At $V_{\rm th}\approx$~300~mV the junction
switches to its low resistance ON state ($R_{\rm ON}\approx0.1$~k$\Omega$). At a subsequent decrease of the bias a linear dependence is observed until the
negative threshold voltage is reached where the OFF state is restored. The slight backward turning of the onset of the OFF to ON switching arises due to the
compensation for the finite, 50--200~$\Omega$ serial resistance of the voltage biasing circuit. By limiting the current in the ON state, this serial resistance
also plays a role in maintaining the stability of the junctions. The observed switching scheme is ideal for memory applications, as the device can be switched
between the two states at a reasonably high bias while $R_{\rm ON}$ and $R_{\rm OFF}$ are optimal for fast readout at low bias.

A similar I-V trace is obtained at 4.2~K as shown in Fig.~\ref{Fig1}(b) testifying to the metallic nature of the system. The linear ON and OFF traces and the
wide range of operation temperature contradict to the thermally activated ionic migration in bulk Ag$_{2}$S \cite{jz900375a}. The observed metallic behavior is
consistent with first principle calculations applicable to a thin surface layer of Ag$_{2}$S deposited on bulk Ag which is expected to undergo a structural
phase transition called spontaneous metallization \cite{nl0711054}. Next we investigate this metallized state experimentally by replacing the PtIr tip by
superconducting Nb and utilizing point contact Andreev reflection (PCAR) spectroscopy \cite{Andreev1964,PhysRevB.25.4515,PhysRevB.54.7366,Soulen1998}.

This method is inspired by the fact that linear conductance measurements alone cannot distinguish between fundamentally different types of junctions exhibiting
the same conductance. The conductance of a nano-scale device is given as $G=2e^2/h\cdot M\cdot T$, where $T$ is the average electron transmission probability
across the device and $M$ is the number of open conductance channels \cite{Landauer_form,nazarovbook}. The latter is approximated as $M\approx(\pi
d/2\lambda_{F})^2$, where $d$ is the device diameter and $\lambda_{F}$ is the Fermi wavelength which is $\approx 0.4\,$nm in bulk Ag. This simplified picture
shows that a k$\Omega$ range device resistance may as well correspond to a large area tunnel junction ($d\gg \lambda_F$) with very small transmission
probability ($T \ll 1$) or, alternatively, to a truly nanometer-scale junction with only a few well transmitting ($T \approx 1$) conductance channels. PCAR
measurements performed on the voltage scale of $\Delta$ can distinguish between these extremities via the nonlinear transport properties of a point contact
connecting a normal metal (N) and a superconductor (S) \cite{Andreev1964}. The transmission can be evaluated by fitting the voltage dependent differential
conductance with the model of Blonder, Tinkham and Klapwijk (BTK) \cite{PhysRevB.25.4515,PhysRevB.54.7366}.

\begin{figure}[t!]
\includegraphics[width=\columnwidth]{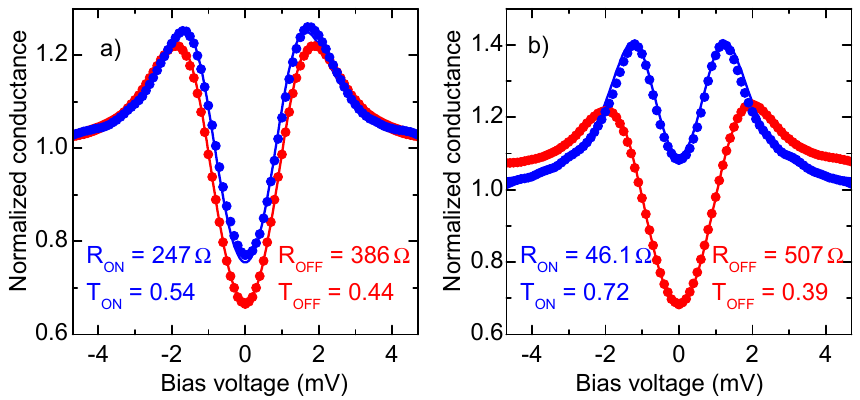}
\caption{(Color online) ON and OFF state finite bias differential conductances obtained from the measured I-V traces by numerical differentiation with respect
to the bias voltage in two representative Ag-Ag$_2$S-Nb junctions at 4.2 K. Each curve is normalized to its high bias $(eV\gg \Delta$) value $G_{\rm ON,OFF}=
\left(R_{\rm ON,OFF} \right)^{-1}$. Continuous lines are fits against the BTK theory resulting in the indicated transmissions.} \label{Fig2}
\end{figure}

Figure~\ref{Fig2} shows the finite bias differential conductance of the ON and OFF states in two representative junctions exhibiting a minor
[Fig.~\ref{Fig2}(a)] and an order of magnitude [Fig.~\ref{Fig2}(b)] change in the normal-state differential conductance $G_{N}$ upon switching. The typical
switching threshold voltages of $V_{\rm th}\approx 300$\,--\,$800$~mV are well separated from the N-S features taking place at the voltage scale of $\Delta$.
The effective transmissions were determined by fitting the differential conductance curves against the modified version of the BTK theory, incorporating a
$\Gamma$ broadening parameter which is necessary to account also for contact diameters comparable to the electronic mean free path
\cite{Plecenik1994,Geresdi2008}. With this extension the spin degenerate model contains five parameters: the temperature, $\Delta$, $G_{N}$, $\Gamma$ and $T$.
A numerically reliable and physically meaningful fitting procedure, however, requires the reduction of such an extended parameter space to a minimum set of key
parameters. After taking extra care of the proper thermal anchoring of the device leads in our $^{4}$He cryostat we fixed the temperature value to 4.2~K.
Finite bias tunneling measurements ($R_{\rm contact}>$100~k$\Omega$) carried out on Ag-Ag$_2$S-Nb samples and on clean Au foils revealed that the bulk value of
$\Delta=$~1.4~meV is a good approximation also in our point contact arrangement. In these data, $\Gamma$ stayed below 5\% of $\Delta$ quantifying the voltage
noise of our setup. After evaluating $G_{N}$ from the high bias linear slopes of the raw I-V traces, the BTK fittings were run with two free parameters $T$ and
$\Gamma$.

In device 1 [Fig.~\ref{Fig2}(a)] $R_{\rm ON}=$~247~$\Omega$ and $R_{\rm OFF}=$~386~$\Omega$, the corresponding transmission probabilities are $0.54$ and
$0.44$, whereas the effective numbers of open conductance channels, estimated as $M=G_{N}/G_{0}T$, are $118$ and $62$, respectively. This shows that both the
ON and OFF states are characterized by rather large transmission values and the 56\% change of the conductance between the two states is equally attributed to
the variations in $M$ and $T$. In device 2 [Fig.~\ref{Fig2}(b)] the order of magnitude change in $G_{N}$ ($G_{\rm ON}/G_{\rm OFF}=11$) is accompanied by a
large change in $M$ ($M_{\rm ON}/M_{\rm OFF}=6.2$) and a minor change in $T$ ($T_{\rm ON}/T_{\rm OFF}=1.8$). The ON and OFF state effective contact diameters
are estimated to be $d_{\rm ON}=2.7\,$nm and $d_{\rm OFF}=2\,$nm in device 1 and $d_{\rm ON}=5\,$nm and $d_{\rm OFF}=2.1\,$nm in device 2, demonstrating that
resistive switching takes place in truly nanometer-scale junctions.

\begin{figure}[t!]
\includegraphics[width=\columnwidth]{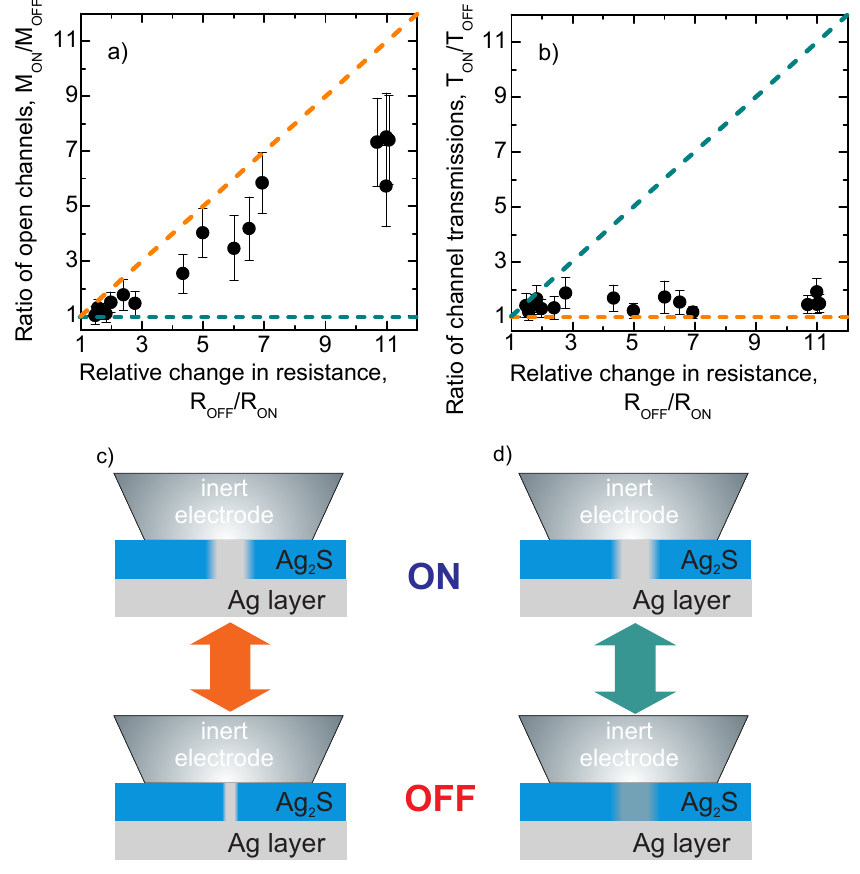}
\caption{(Color online) Analysis of the finite bias differential conductance data obtained in various Ag-Ag$_2$S-Nb devices. The ratios of the number of
conducting channels (a) and the average transmission probabilities (b) in the ON and OFF states of each junction are shown as a function of the corresponding
relative change in the normal resistance of the devices. The orange dash line displays the limiting case where only $M$ changes as illustrated in (c). The
green dash line indicates the opposite scenario when the change in the resistance is solely attributed to the variation of $T$ as illustrated in (d). In the
lower panels the width and opacity of the grey region across the Ag$_{2}$S layer (blue) represent the number and transmission of the conducting channels,
respectively.} \label{Fig3}
\end{figure}

In order to verify the statistical relevance of these findings, $T$ and $d$ were evaluated for various junctions yielding to $T_{\rm ON}=0.62\pm0.1$  and
$T_{\rm OFF}=0.42\pm0.07$ at effective junction diameters of $d=$~2--5~nm. Figure~\ref{Fig3} shows the relative changes in $T$ and $M$ upon switching as a
function of the resistance ratio of the corresponding ON and OFF states. The two limiting cases of the unchanged transmissions and unchanged channel numbers
are indicated in Figs.~\ref{Fig3}(a) and \ref{Fig3}(b) by the orange and green dash lines and are also schematically illustrated in Figs.~\ref{Fig3}(c) and
\ref{Fig3}(d), respectively. The numerical accuracy of $T$ is 20\% in the OFF states and better than 5\% in the ON states as explained in the Appendix and
indicated by the error bars in Figs.~\ref{Fig3}(a) and \ref{Fig3}(b). In the studied junctions with ON and OFF state resistances of 50--1000~$\Omega$, $\Gamma$
is ranging 50--15\% of $\Delta$, respectively. These values are significantly higher than those obtained in the tunneling regime in agreement with previous
studies carried out on various diffusive systems \cite{Geresdi2008} indicating that in spite of the small junction diameters electron transport is not entirely
ballistic due to the rather short mean free path of 1.8~nm in Ag$_{2}$S \cite{Zemek2001}.

\begin{figure}[t!]
\includegraphics[width=\columnwidth]{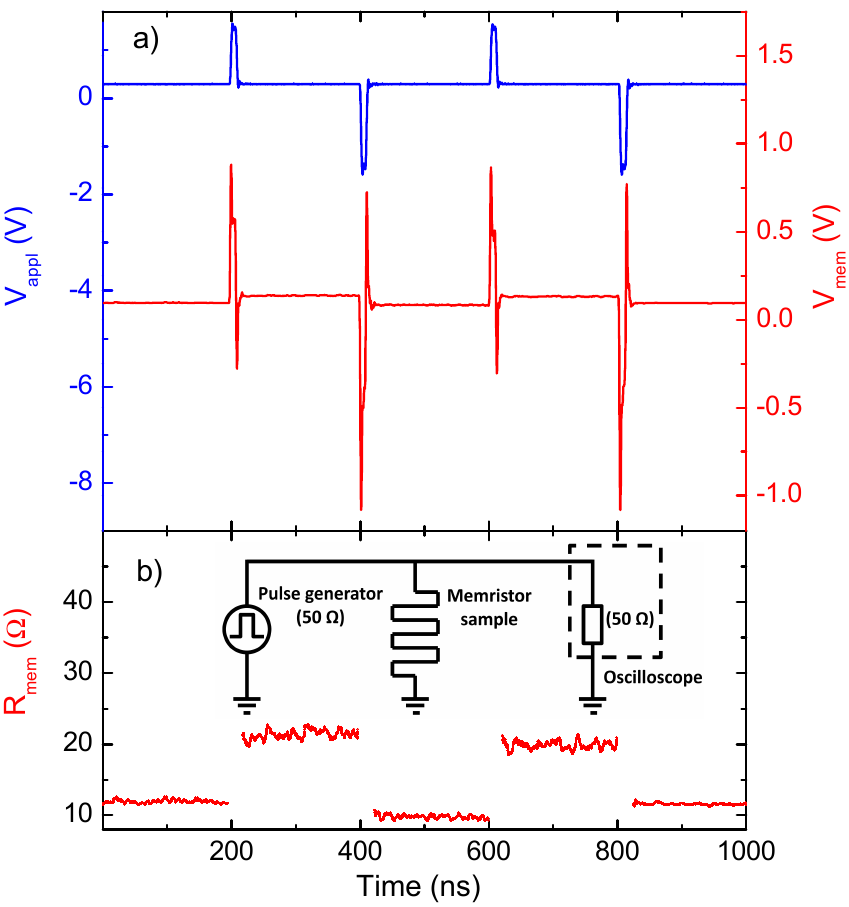}
\caption{(Color online) (a) Alternating 10 ns wide voltage pulses ($V_{\rm appl}$, blue line) recorded in an open circuit configuration and measured on a
Ag-Ag$_2$S-PtIr junction ($V_{\rm mem}$, red line) at room temperature. During the 200 ns time between the pulses a 350 mV dc read-out voltage was applied. (b)
The calculated resistance demonstrates reproducible switching behavior. Data acquired within the pulse duration is not shown. The inset illustrates the
equivalent circuit of the setup.} \label{Fig4}
\end{figure}

Our analysis shows that (i) regardless of the relative change in their resistances, all devices exhibit high transmission probabilities in both states
corresponding to metallic conductances and (ii) their conductance changes can rather be attributed to a variation in $M$ while $T$ is less significantly
changed during the switching. These results give clear evidence that the metallic channels are not completely destructed when the device is switched off, only
their effective cross section is reduced while the highly transmitting nature is preserved. This is in good agreement with the presence of spontaneously
metallized Ag$_{2}$S channels \cite{nl0711054} and is in contrast with the simplified picture of the formation and complete destruction of pure Ag
nanofilaments. The above characteristics make Ag$_{2}$S based devices particularly suitable to build resistive memory elements with technologically ideal
resistances of $\leq$1~k$\Omega$ and diameters well below the resolution of nowadays lithographic techniques. Note that tunnel junctions of similar sizes would
yield to orders of magnitude larger resistances challenging their fast electronic applications.

Finally we demonstrate that the resistive switching of the investigated nanometer-scale metallic channels is a fast process taking place on a nanosecond time
scale. Real-time pulsed measurements were performed in vacuum conditions at room temperature by connecting the Ag-Ag$_2$S-PtIr junction to an arbitrary
waveform generator and a digital oscilloscope of 2~ns time resolution via a transmission line in a circuit schematically illustrated in the inset of
Fig.~\ref{Fig4}(b). Subsequent switchings between the ON and OFF states were achieved by voltage pulses of 10 ns duration, 1.5 V amplitude and alternating
sign. Between the pulses the device resistance was monitored by applying a 350 mV dc bias. The blue line in Fig.~\ref{Fig4}(a) shows the applied pulse train in
an open circuit configuration when the sample is not connected and the transmission line is terminated by an open end resulting in the total reflection of the
pulses. In comparison, the red line corresponds to the situation when the transmitted signal from the pulse generator to the scope is superimposed on a
partially reflected signal emerging due to the finite resistance of the memristor device. Figure~\ref{Fig4}(b) shows the calculated resistances of $R_{\rm
OFF}$=20~$\Omega$ and $R_{\rm ON}$=12~$\Omega$. The subsequent, reproducible resistance plateaus are well separated both from each other and from the noise
floor of the measurement, satisfying the two most profound requirements of memory applications. While fast pulsing measurements favor contact impedances close
to 50~$\Omega$, fast switching at higher contact resistances were also observed, though with lower resolution.

\begin{figure}[t!]
\includegraphics[width=0.9\columnwidth]{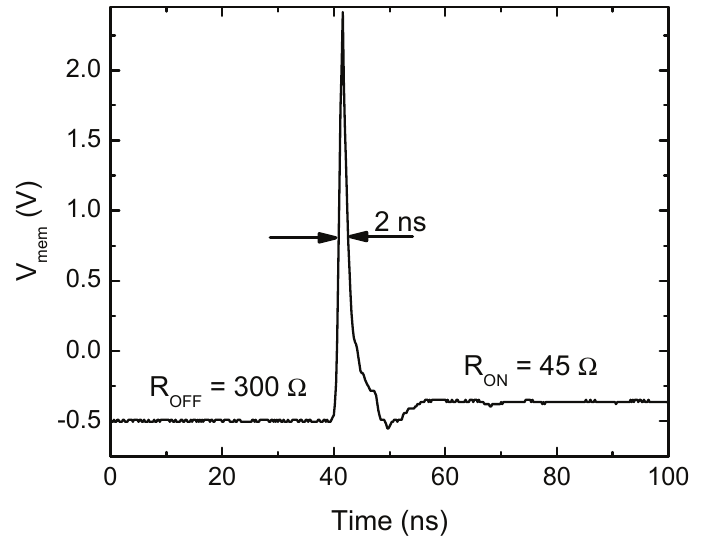}
\caption{OFF to ON switching due to a single voltage pulse of 500~ps width and 2.5~V amplitude. The calculated resistance values are indicated. The apparent
2~ns pulse width arises due to the limited detector bandwidth.} \label{Fig5}
\end{figure}

While the $\geq$1.5 ns rise/fall times of our arbitrary waveform generator limit the investigation of shorter time domains by voltage pulses of alternating
sign, the single shot experiment shown in Fig.~\ref{Fig5} demonstrates that resistive switching in Ag$_{2}$S can also be induced by voltage pulses as short as
500 ps. Such measurements were carried out by using a custom built rise time avalanche pulse generator specified to provide unipolar voltage pulses of 500 ps
width and $\leq$10 V amplitude \cite{pulsegen}. The low bias readout between the short pulses and the subsequent re-initialization of the OFF state were
performed by a variable dc offset added to the biasing circuit as explained in the Appendix. Whereas the real-time response of the device is not fully resolved
in the experiment due to instrumental and circuit limitations, it is evident that a sub-nanosecond pulse can initiate a resistive switching.

In conclusion, we studied the resistive switching of Ag-Ag$_{2}$S-Me memristive nanojunction devices. We showed that by suitable sample preparation
reproducible resistive switching and readout can be performed where \emph{both} the ON and OFF states are metallic, characterized by technologically optimal
$\leq$1~k$\Omega$ resistances and similar device functionalities down to cryogenic temperatures. We introduced PCAR spectroscopy to determine the size and
transmission probabilities of the active volume of the devices which revealed a small number of highly transmitting nanoscale conducting channels with reduced
but not completely dissolved junction area also in the OFF state. The relatively low resistance ON and OFF states enable fast operation at room temperature as
demonstrated by the nanosecond switching times. Our results suggest that Ag$_{2}$S represents a promising material basis for a future generation of high speed
resistive switching memory devices overriding the downscaling limitations of nowadays CMOS technology.

Enlightening discussions with \'Ad\'am Sz\"ull\H{o} and his circuit designing contribution to the fast pulsing experiments are acknowledged. This work has been
supported by the Hungarian research funds OTKA CNK80991, K105735, T\'AMOP 4.2.1.B-09/1--2010-0002. M.C. is a grantee of the Bolyai J\'anos Research Scholarship
of the HAS and acknowledges financial support of the European Union 7th Framework Programme (Grant No.~293797).

\section*{APPENDIX}

Ag thin films with a nominal thickness of 80 nm were vacuum evaporated onto a Si substrate. The thin Ag$_2$S layers were grown by depositing sulfur onto the Ag
surfaces in a clean environment. First, analytic grade sulfur powder was loaded in a quartz tube, melted and cooled back in order to ensure a homogenous
source. The thin film sample was then loaded in the tube to a distance of $2$ cm from the sulfur. After loading both the sulfur and the sample, the tube was
evacuated to $10^{-5}$ mbar pressure. Then the temperature was ramped up to 60 $^{\circ}$C and the sublimation of the sulfur was performed in a static vacuum
for 2--10 minutes. Finally, the temperature was rapidly ramped down.

The samples were characterized by He-RBS (Rutherford Backscattering Spectrometry) and ERDA (Elastic Recoil Detection Analysis) \cite{Kotai1994588}. They
exhibit inhomogeneous sulfur concentration profiles consistent with the presence of an Ag$_2$S surface layer \cite{GeresdiMRS2011}.

\renewcommand\thefigure{A\arabic{figure}}
\setcounter{figure}{0}

\begin{figure}[b!]
\includegraphics[width=\columnwidth]{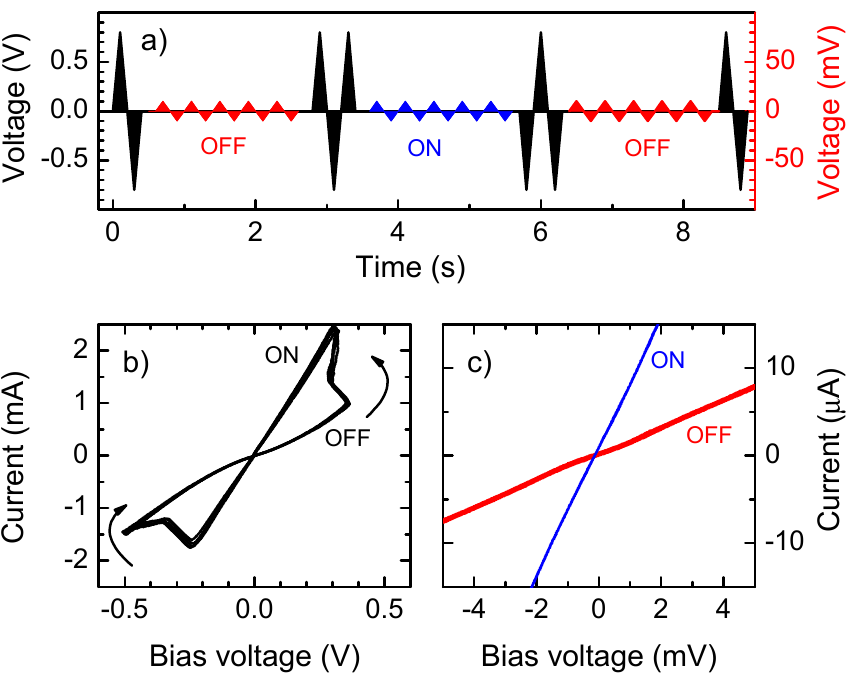}
\caption{(a) Biasing scheme of the I-V measurements. The 800 mV amplitude voltage sweeps (black triangles) are used to record the hysteretic switching
characteristics and to prepare the subsequent OFF/ON/OFF states for the voltage sweeps performed repeatedly on the scale of the Nb superconducting gap between
$\pm$10 mV (red and blue triangles). (b) Hysteretic high bias I-V characteristics of an Ag-Ag$_2$S sample and a superconducting Nb tip. The overlap of the 4
subsequent loops demonstrate reproducible resistive switching over the entire pulsing sequence. The arrows indicate the direction of the bias sweep. (c) Low
bias I-V traces corresponding to the ON (1 trace) and OFF states (2 overlapping traces). All data were taken at 4.2 K.} \label{FigS1}
\end{figure}

\begin{figure}[t!]
\includegraphics[width=\columnwidth]{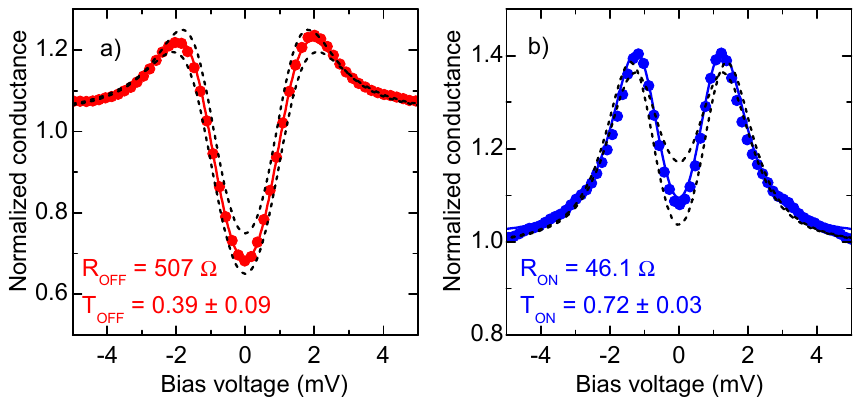}
\caption{Representative finite bias conductance in the OFF (a) and ON states (b) of an Ag-Ag$_{2}$S-Nb junction at 4.2 K obtained by numerical differentiation
of the measured low bias I-V traces. Continuous lines denote fits using the BTK theory with the obtained transmission values shown in the graphs. Note that
each curve is normalized to its high bias $(eV\gg \Delta$ = 1.4 meV) value $G_{\rm ON,OFF}= \left(R_{\rm ON,OFF} \right)^{-1}$. The dashed lines correspond to
the best fits using the detuned transmission values as labeled in the graphs testifying to the accuracy of our analysis on $T$.} \label{FigS2}
\end{figure}

\begin{figure}[t!]
\includegraphics[width=\columnwidth]{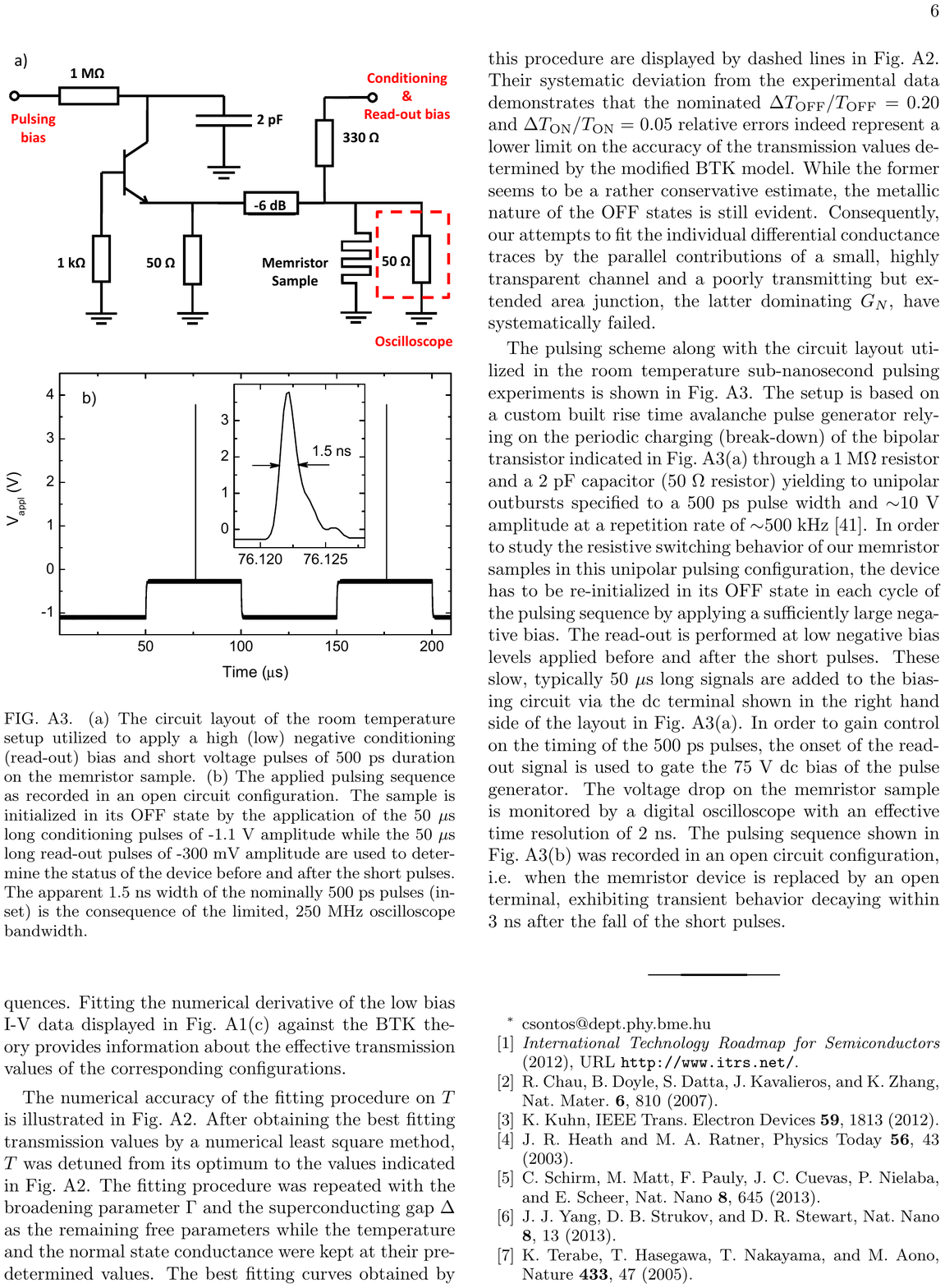}
\caption{(a) The circuit layout of the room temperature setup utilized to apply a high (low) negative conditioning (read-out) bias and short voltage pulses of
500 ps duration on the memristor sample. (b) The applied pulsing sequence as recorded in an open circuit configuration. The sample is initialized in its OFF
state by the application of the 50 $\mu$s long conditioning pulses of -1.1 V amplitude while the 50 $\mu$s long read-out pulses of -300 mV amplitude are used
to determine the status of the device before and after the short pulses. The apparent 1.5 ns width of the nominally 500 ps pulses (inset) is the consequence of
the limited, 250 MHz oscilloscope bandwidth.} \label{FigS3}
\end{figure}

Nanoscale contacts were created by gently touching the sample surface with a mechanically sharpened PtIr or Nb tip. For coarse adjustment a screw thread
mechanism was used, whereas for the fine positioning a three dimensional piezo scanner was applied. Using this technique numerous contacts were created with
reproducible I-V characteristics. The bias voltage was applied to the junctions by utilizing the analog outputs of a National Instruments data acquisition
card. The output voltage was divided and filtered in order to ensure low noise on the contact. The current was measured using a variable range I-V converter
and processed by the data acquisition card. The current-voltage characteristics of the junctions were recorded during repeated bias voltage sweeps.

The bias scheme of the I-V measurements carried out with a superconducting Nb tip at 4.2 K is shown in Fig.~\ref{FigS1}(a). The subsequent resistive switchings
are obtained by voltage sweeps of 800 mV amplitude and alternating sign while the ON and OFF state resistances as well as the non-linear current contributions
induced by the superconducting tip are detected by low amplitude ($\pm$10 mV) triangular voltage pulses. The first high bias sweep is performed in order to
verify the switching and to initialize the device in its OFF state. This is followed by the acquisition of the low bias IV data repeated 5 times for averaging
purposes in order to further improve the signal to noise ratio. The next high bias sweep covers one and a half of the total hysteresis loop preparing the
device in its ON state before the second sequence of the low bias measurements probing the ON state transport properties are executed. In order to confirm that
upon the first OFF to ON switching and ON state measurements the device preserved its reversible configuration, we restored and tested the OFF state again by
applying another high bias loop and a low bias sequence, respectively. The reproducibility of the hysteretic high bias I-V traces shown in Fig.~\ref{FigS1}(b)
demonstrate that the two dominant device configurations determining the ON and OFF states did not change over the above sweeping sequences. Fitting the
numerical derivative of the low bias I-V data displayed in Fig.~\ref{FigS1}(c) against the BTK theory provides information about the effective transmission
values of the corresponding configurations.

The numerical accuracy of the fitting procedure on $T$ is illustrated in Fig.~\ref{FigS2}. After obtaining the best fitting transmission values by a numerical
least square method, $T$ was detuned from its optimum to the values indicated in Fig.~\ref{FigS2}. The fitting procedure was repeated with the broadening
parameter $\Gamma$ and the superconducting gap $\Delta$ as the remaining free parameters while the temperature and the normal state conductance were kept at
their predetermined values. The best fitting curves obtained by this procedure are displayed by dashed lines in Fig.~\ref{FigS2}. Their systematic deviation
from the experimental data demonstrates that the nominated $\Delta T_{\rm OFF}/T_{\rm OFF}=$~0.20 and $\Delta T_{\rm ON}/T_{\rm ON}=$~0.05 relative errors
indeed represent a lower limit on the accuracy of the transmission values determined by the modified BTK model. While the former seems to be a rather
conservative estimate, the metallic nature of the OFF states is still evident. Consequently, our attempts to fit the individual differential conductance traces
by the parallel contributions of a small, highly transparent channel and a poorly transmitting but extended area junction, the latter dominating $G_{N}$, have
systematically failed.

The pulsing scheme along with the circuit layout utilized in the room temperature sub-nanosecond pulsing experiments is shown in Fig.~\ref{FigS3}. The setup is
based on a custom built rise time avalanche pulse generator relying on the periodic charging (break-down) of the bipolar transistor indicated in
Fig.~\ref{FigS3}(a) through a 1 M$\Omega$ resistor and a 2 pF capacitor (50 $\Omega$ resistor) yielding to unipolar outbursts specified to a 500 ps pulse width
and $\sim$10 V amplitude at a repetition rate of $\sim$500 kHz \cite{pulsegen}. In order to study the resistive switching behavior of our memristor samples in
this unipolar pulsing configuration, the device has to be re-initialized in its OFF state in each cycle of the pulsing sequence by applying a sufficiently
large negative bias. The read-out is performed at low negative bias levels applied before and after the short pulses. These slow, typically 50 $\mu$s long
signals are added to the biasing circuit via the dc terminal shown in the right hand side of the layout in Fig.~\ref{FigS3}(a). In order to gain control on the
timing of the 500 ps pulses, the onset of the read-out signal is used to gate the 75 V dc bias of the pulse generator. The voltage drop on the memristor sample
is monitored by a digital oscilloscope with an effective time resolution of 2 ns. The pulsing sequence shown in Fig.~\ref{FigS3}(b) was recorded in an open
circuit configuration, i.e. when the memristor device is replaced by an open terminal, exhibiting transient behavior decaying within 3 ns after the fall of the
short pulses.


\end{document}